\pdfoutput=1

\documentclass[11pt]{article}

\usepackage[final]{acl}

\usepackage{times}
\usepackage{latexsym}
\usepackage{multirow}
\usepackage{float}
\usepackage{booktabs}
\usepackage{amsmath}
\usepackage{enumitem}
\usepackage[T1]{fontenc}

\usepackage[utf8]{inputenc}

\usepackage{microtype}

\usepackage{inconsolata}

\usepackage{graphicx}

\title{ArabEmoNet: A Lightweight Hybrid 2D CNN-BiLSTM Model with Attention for Robust Arabic Speech Emotion Recognition}

\author{Ali Abouzeid\thanks{Equal contribution}, Bilal Elbouardi\footnotemark[1], Mohamed Maged\footnotemark[1], Shady Shehata  \\
  Mohamed bin Zayed University of Artificial Intelligence, University of Waterloo \\
   \texttt{ali.abouzeid, Bilal.ElBouardi, Mohamed.Elsetohy}@mbzuai.ac.ae \\
   \texttt{shady.shehata@uwaterloo.ca}
  }

\begin{document}

\maketitle
\begin{abstract}
Speech emotion recognition is vital for human-computer interaction, particularly for low-resource languages like Arabic, which face challenges due to limited data and research. We introduce ArabEmoNet, a lightweight architecture designed to overcome these limitations and deliver state-of-the-art performance. Unlike previous systems relying on discrete MFCC features and 1D convolutions, which miss nuanced spectro-temporal patterns, ArabEmoNet uses Mel spectrograms processed through 2D convolutions, preserving critical emotional cues often lost in traditional methods. While recent models favor large-scale architectures with millions of parameters, ArabEmoNet achieves superior results with just 1 million parameters, which is 90 times smaller than HuBERT base and 74 times smaller than Whisper. This efficiency makes it ideal for resource-constrained environments. ArabEmoNet advances Arabic speech emotion recognition, offering exceptional performance and accessibility for real-world applications.
\end{abstract}

\section{Introduction}

\label{s:intro}

Speech Emotion Recognition (SER) is essential for improving human-computer interaction, particularly in linguistically diverse contexts like Arabic speech. The complexity of detecting emotions from speech arises from variations in prosody, phonetics, and speaker expression. Over time, SER has evolved from statistical approaches to deep learning, significantly enhancing recognition accuracy.

Early SER systems relied on handcrafted acoustic features (e.g., pitch, energy, and MFCCs) processed using classical machine learning models like Support Vector Machines (SVMs) and Gaussian Mixture Models (GMMs) \cite{lieskovska2021review}. While effective, these methods struggled with cross-dataset generalization, particularly in Arabic speech, which exhibits rich phonetic and prosodic diversity. Deep learning mitigated these limitations by enabling automatic feature extraction, with CNNs capturing localized spectro-temporal patterns and LSTMs modeling sequential dependencies \cite{fayek2017evaluating}. However, many Arabic SER systems still rely on MFCCs and 1D convolutions, which fail to capture essential spectral-temporal structures for robust emotion recognition.

Transformer-based models \cite{vaswani2017attention} introduced attention mechanisms to dynamically focus on emotionally salient speech segments \cite{mirsamadi2017automatic}. While effective in modeling long-range dependencies and parallelizing computations across emotional speech sequences, their high computational complexity ($O(n^2)$ for self-attention) and substantial memory requirements render them impractical for resource-constrained environments. To address these constraints, we propose ArabEmoNet, a lightweight architecture leveraging Mel spectrograms with 2D convolutions, effectively capturing both fine-grained spectral features and global contextual relationships \cite{8282315}.

Our model achieves competitive accuracy with just 0.97M parameters, making it significantly more efficient than HuBERT \cite{hsu2021hubert} and Whisper \cite{radford2022whisper} while maintaining state-of-the-art performance. Additionally, we augmented the data by integrating SpecAugment \cite{park2019specaugment} and Additive White Gaussian Noise (AWGN), which enhances the robustness of our model \cite{huh2024comparisonspeechdataaugmentation}.

Experiments on KSUEmotions \cite{meftah2021ksuemotion} and KEDAS \cite{belhadj2022kedas} datasets confirm that ArabEmoNet surpasses prior architectures while maintaining efficiency, marking a significant step forward in Arabic SER.

The main contributions of this paper can be summarized as follows:
\begin{itemize}
    \item We propose ArabEmoNet: a novel lightweight hybrid architecture combining 2D Convolutional Neural Networks (CNN) with Bidirectional Long Short-Term Memory (BiLSTM) and an attention mechanism 
    \item ArabEmoNet (1M parameters) achieves superior results with just 1 million parameters---90 times smaller than HuBERT base (95M parameters)  and 74 times smaller than Whisper (74M parameters). 
    \item We demonstrate ArabEmoNet's superior performance by achieving state-of-the-art results on the KSUEmotion and KEDAS datasets, surpassing previous benchmark models. 
    
\end{itemize}

\begin{figure*}[h!] 
\centering
\includegraphics[width=1.0\textwidth]{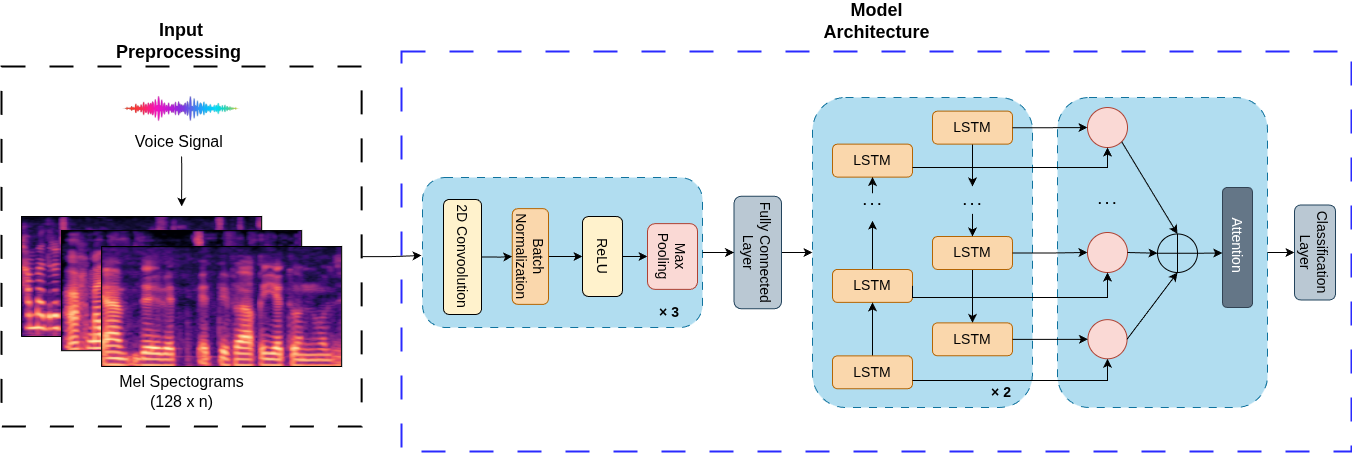}
\caption{ArabEmoNet:2D CNN-Attention and BiLSTM Model Architecture.}
\label{fig:model_arch}
\end{figure*}
\section{Related Work}

\label{s:related-work}

Speech Emotion Recognition (SER) has been an active area of research for decades. Traditional approaches often relied on statistical evaluations of speech features such as pitch, energy, and spectral coefficients, combined with classifiers such as SVMs or Hidden Markov Models (HMMs) \cite{iliou2009statistical} and \cite{nwe2003speech}. Although these methods provided foundational information, they struggled to generalize across different datasets and languages.

The advent of deep learning significantly advanced SER by enabling the extraction of hierarchical feature representations. Convolutional Neural Networks (CNNs) demonstrated their ability to capture local spectral features, while Recurrent Neural Networks (RNNs), including Long Short-Term Memory (LSTM) networks, excelled at modeling temporal dependencies \cite{sainath2015convolutional} and \cite{trigeorgis2016adieu}. End-to-end frameworks combining CNNs and RNNs have further improved performance by learning from raw or minimally processed data \cite{fayek2017evaluating}.

Attention mechanisms, introduced in \cite{bahdanau2015neural} and further refined in \cite{vaswani2017attention}, have recently shown promise in SER. They allow models to focus on the most relevant parts of the input, which is particularly useful for tasks such as emotion recognition, where only a subset of characteristics may contribute to emotional cues \cite{mirsamadi2017automatic} and \cite{neumann2017attentive}.

Previous work in  \cite{hifny2019efficient} proposed an efficient Arabic Speech Emotion Recognition (SER) system, combining Convolutional Neural Networks (CNNs), Bidirectional LSTMs (BiLSTMs), and attention mechanisms, which achieved state-of-the-art performance on the KSUEmotions dataset. However, their approach relied on 13-feature Mel Frequency Cepstral Coefficients (MFCCs) and 1D convolutions, which may restrict the richness of captured acoustic features. Building on the insights from their work and the demonstrated importance of leveraging richer representations by \cite{meng2019speech} and \cite{lieskovska2021review}, we diverge from \cite{hifny2019efficient} by utilizing Log-Mel spectrograms as input, which offer a more comprehensive acoustic feature representation, and employs 2D convolutions to effectively capture spatial dependencies within these spectrograms

Furthermore, advances in data augmentation techniques, such as SpecAugment \cite{park2019specaugment} and Gaussian noise, have addressed data scarcity challenges, enhancing the generalization capabilities of SER systems. Studies have also investigated the impact of kernel sizes in CNN layers and regularization techniques to optimize performance \cite{kumar2024logmel}.

\section{Proposed Approach}
\label{s:prob-approach}

In this work, we introduce ArabEmoNet, a dedicated 2D NN-Attention and BiLSTM framework optimized for Arabic Speech Emotion Recognition. Our model processes Log-Mel spectrograms to effectively capture the multifaceted nature of emotional speech through three complementary components: 2D convolutional layers that identify emotion-specific spectral patterns, bidirectional LSTMs that model the temporal evolution of emotional cues, and an attention mechanism that highlights emotionally salient segments within utterances. This integrated approach addresses the unique challenges of recognizing Arabic emotional expressions while maintaining a lightweight, efficient architecture. Figure \ref{fig:model_arch} illustrates our complete model design.

\subsection{Input Prepossessing} 

For our classification model, raw audio signals are transformed into Log-Mel spectrograms. This process involves computing the Mel spectrogram using a Fast Fourier Transform (FFT) window length of 2048 samples and a hop length of 256 samples. We generate 128 Mel bands across a frequency range from 80 Hz to 7600 Hz . A Hann window is applied to each frame to minimize spectral leakage. Subsequently, the resulting Mel spectrogram is converted to a logarithmic scale (decibels), referenced to the maximum power, to optimize the dynamic range for neural network processing.

\subsection{Data Augmentation}
To improve the generalization ability of the model and mitigate overfitting, we incorporate Gaussian noise augmentation during training. This technique simulates variations in the input data and leads to a more robust model. Optimization is performed using the Adam optimizer, which adapts learning rates for each parameter based on the first and second moments of the gradients. Additionally, we utilize batch normalization and early stopping based on validation loss to further stabilize the training process and prevent overfitting.\\

\subsection{Feature Extraction via Convolutional Layers} The initial stage of the model employs a series of convolutional layers to extract high-level representations from the input Mel spectrograms. These layers are responsible for detecting local time-frequency patterns that are crucial for emotion discrimination. Mathematically, the feature maps $\mathbf{F}_l$ at layer $l$ are computed as:
\[
\mathbf{F}_{l} = \sigma\left( \text{Conv2D}(\mathbf{F}_{l-1}, \mathbf{W}_{l}, \text{padding}=p_{l}) + b_{l} \right)\]
where $\mathbf{F}_{l-1}$ represents the input to the current layer (with the initial input being the spectrogram $\mathbf{S}$), $\mathbf{W}_{l}$ and $b_{l}$ denote the learnable weights and biases, respectively, $p_{l}$ is the specified padding, and $\sigma$ is the ReLU activation function. It is important to note that we employ 2D CNNs rather than 1D CNNs because Mel spectrograms provide a two-dimensional (time-frequency) representation. This allows the model to capture both temporal and spectral dependencies more effectively. The use of multiple convolutional layers, combined with max-pooling and dropout, enhances the network's ability to learn robust, hierarchical feature representations while mitigating overfitting.
Following the convolutional layers, the extracted features are passed through a fully connected layer before being passed to the next stage. 

\subsection{Temporal Modeling with Bidirectional LSTM}
After the convolutional layers, the network integrates a Bidirectional LSTM to model the temporal structure and contextual dependencies across time frames. By processing the sequential output in both forward and backward directions, the BiLSTM effectively captures transitions between emotional states, ensuring a more nuanced understanding of temporal variations in speech. The hidden state at time step \(t\) is given by:

\[
\mathbf{h}_t = \left[ \overrightarrow{\mathbf{h}}_t; \overleftarrow{\mathbf{h}}_t \right],
\]

where \(\overrightarrow{\mathbf{h}}_t\) and \(\overleftarrow{\mathbf{h}}_t\) denote the forward and backward hidden states, respectively. This bidirectional processing is particularly important for SER tasks, as emotions in speech often evolve gradually rather than appearing in isolation. Capturing the transitions between emotional states allows the model to account for contextual cues, such as shifts in pitch, intensity, and rhythm, which are crucial for accurately interpreting emotional expressions over time.

\subsection{Attention Mechanism} 
To enhance the model's ability to distinguish subtle variations in emotional expressions, an attention mechanism is integrated atop the BiLSTM outputs. This mechanism computes a context vector \(\mathbf{c}\) that selectively aggregates the BiLSTM hidden states, assigning higher importance to frames that carry more salient emotional cues, thereby improving emotion classification. The context vector is defined as:

\[
\mathbf{c} = \sum_{t} \alpha_t \mathbf{h}_t, \quad \text{with} \quad \alpha_t = \frac{\exp(e_t)}{\sum_{k} \exp(e_k)},
\]

where the attention score \(e_t\) is computed as:

\[
e_t = \tanh\left(\mathbf{w}_e^\top \mathbf{h}_t + b_e\right).
\]

Here, \(\mathbf{w}_e\) and \(b_e\) are learnable parameters that transform the hidden states into a scalar score, and the softmax function normalizes these scores into a probability distribution over time steps. By dynamically focusing on the most emotionally informative segments of the speech signal, this mechanism enhances the model's ability to capture key variations in tone, prosody, and intensity that define different emotional states, making it more effective for Speech Emotion Recognition (SER).

\vspace{0.2cm}
\subsection{Classification Layer:} Finally, the context vector is passed through one fully connected layer, culminating in an output layer that produces the logits corresponding to the target emotion classes:
\[
\mathbf{o} = \mathbf{W}_o \mathbf{c} + b_o.
\]
The logits are then typically passed through a softmax function during training to compute the cross-entropy loss for classification. The entire architecture is illustrated in Figure~\ref{fig:model_arch}.

\begin{table}[t]
  \centering
  
  \resizebox{0.48\textwidth}{!}{%
    \begin{tabular}{@{}ll@{}}
      \toprule
      \textbf{Component}   & \textbf{Configuration} \\ \midrule
      Convolutional Layers & 3 stages with filters: 32, 64, 128 \\
                           & Kernel: $7\times7$, ReLU activation \\
                           & Max pooling: $2\times2$, dropout: 0.3 \\ \midrule
      Fully Connected      & Input: $128 \times H'$; Output: 128 \\
                           & ReLU activation; dropout: 0.3 \\ \midrule
      BiLSTM               & 2 layers, 64 hidden units per direction \\
                           & Dropout: 0.3 \\ \midrule
      Attention            & Applied to 128-dim BiLSTM output \\ \midrule
      Classification       & Units equal to number of emotion categories \\
      \bottomrule
    \end{tabular}%
  }
  \caption{Model Hyperparameter Configuration}
  \label{tab:config}
\end{table}

\section{Experimental setup}
\label{s:experimental_setup}

\subsection{Training Platform}
Training was done on a single Nvidia RTX 4090 GPU with 24 GB of memory. The training process utilized the Adam optimizer with an initial learning rate of $1 \times 10^{-4}$ and a weight decay of $1 \times 10^{-5}$. An adaptive learning rate scheduler that reduces the learning rate when a metric's improvement plateaus was incorporated to adjust the learning rate during training, and the Adam optimizer was included. 

\subsection{Baselines}
For our baseline models, we used Whisper-base, Whisper-small, and HuBERT-base speech encoders due to their vast popularity in the speech domain. We applied two identical feed-forward sublayers, each comprising a fully connected layer followed by a ReLU activation function and a dropout layer. This feed-forward block is repeated twice. After the feed-forward modules, the output is passed to a final classification layer that maps the learned features to the desired output classes.
We trained the models using Adam optimizer with learning rate $1 \times 10^{-3}$ and dropout 0.5. 
In addition to these general speech encoders, we also compared ArabEmoNet against several dataset-specific baseline models from the literature:
\begin{itemize}
    \item For the KSUEmotion dataset, we compared against the ResNet-based Architecture \cite{meftah2021ksuemotion} and the CNN-BLSTM-DNN Model \cite{hifny2019efficient}. 
    \item For the KEDAS dataset, baseline \cite{belhadj2022kedas} reported in the original dataset paper.

\end{itemize}

\subsection{Datasets}

In this work, we utilized two Arabic emotional speech datasets: the KSUEmotions corpus and KEDAS, both designed to advance speech emotion recognition (SER) research in Arabic, addressing the scarcity of non-English SER resources. We sampled both datasets at their native frequencies: 16kHz for KSUEmotions and 48kHz for KEDAS. To handle varying sequence lengths in the dataset, shorter sequences within a batch were padded with zeros to match the longest sequence. 

\subsubsection{KSUEmotions Dataset}
The KSUEmotions corpus~\cite{meftah2021ksuemotion} provides recordings from 23 native Arabic speakers (10 males, 13 females) representing diverse dialectal backgrounds from Yemen, Saudi Arabia, and Syria. The corpus was collected in two phases:

\begin{enumerate}[label=\arabic*)]
    \item Phase 1: Included 20 speakers (10 males, 10 females) recording five emotions: neutral, sadness, happiness, surprise, and questioning, totaling 2 hours and 55 minutes of high-quality audio recorded in controlled environments.
    
    \item Phase 2: Featured 14 speakers (7 males and 4 females from Phase 1, plus 3 new Yemeni females), replacing the questioning emotion with anger, contributing an additional 2 hours and 15 minutes of recordings.
\end{enumerate}

\subsubsection{KEDAS Dataset}
The KEDAS dataset \cite{belhadj2022kedas} comprises 5000 audio recording files in standard Arabic, featuring five emotional states: anger, happiness, sadness, fear, and neutrality. The recordings were collected from 500 actors within the university community, including students, professors, and staff. The dataset is based on 10 carefully selected phrases commonly used in communication, chosen through literary and scientific studies. The data collection and validation process involved 55 evaluators, including Arabic linguists, literary researchers, and clinical psychology specialists, ensuring high-quality emotional content and linguistic accuracy.

\begin{table*}
\resizebox{1.0\textwidth}{!}
{
\begin{tabular}{l|l|p{1.5cm}p{1.5cm}p{1.5cm}p{1.5cm}}
\toprule
\textbf{Dataset} & \textbf{Model} & \textbf{Accuracy (\%) $\uparrow$} & \textbf{Micro F1 (\%)} & \textbf{Macro F1 (\%)} & \textbf{Params (M)} \\
\midrule
\multirow{8}{*}{KSUEmotion}
& Whisper-base \cite{radford2022whisper} & 78.81 & 76.77 & 78.81 & 74 \\
& Hubert-base-Emotion & 84.30 & 83.00 & 84.00 & 95 \\
& ResNet-based Architecture \cite{meftah2021ksuemotion} & 85.53 & 85.53 & 85.53 & 25 \\
& Whisper-small \cite{radford2022whisper}& 85.98 & 85.96 & 85.98 & 244 \\
& Hubert-base \cite{hsu2021hubert} & 87.04 & \underline{87.22} & 87.04 & 95 \\
& ArabEmoNet (Transformer) - \textbf{Ours} & 86.66 & 86.66 & 86.66 & 1 \\
& CNN-BLSTM-DNN Model \cite{hifny2019efficient} & \underline{87.20} & 87.20 & \underline{87.20} & - \\

& ArabEmoNet - \textbf{Ours} & \textbf{91.48} & \textbf{91.48} & \textbf{91.46} & 1 \\
\midrule
\multirow{2}{*}{KEDAS}
& Baseline Model \cite{belhadj2022kedas} & 75.00 & 75.00 & 75.00 & - \\
& Whisper-base \cite{radford2022whisper} & 97.60 & 97.56 & 97.60 & 74 \\
& Hubert-base-Emotion & 98.00 & 97.98 & 98.00 & 95 \\
& Hubert-base \cite{hsu2021hubert} & 99.35 & \textbf{99.48} & \textbf{99.50} & 95 \\

& Whisper-small \cite{radford2022whisper} & \underline{99.40} & 99.38 & 99.40 & 244 \\

& ArabEmoNet - \textbf{Ours} & \textbf{99.46} & \underline{99.46} & 
\underline{99.42} & 1 \\
\bottomrule
\end{tabular}
}
\caption{Comparison of Models on KSUEmotion and KEDAS Datasets}
\label{tab:combined_model_comparison}
\end{table*}

\subsection{Evaluation}

To evaluate our classification model's performance, we used two key metrics: Macro F1-score and Micro F1-score. Since no specific train-test split was provided for the datasets, we follow \cite{hifny2019efficient} and report the average of a 5-fold cross-validation with stratified splits on both datasets.

\subsubsection{Macro F1-Score}
The macro F1-score \cite{sokolova2009systematic} calculates the unweighted mean of F1-scores for each class. It treats all classes equally, regardless of their size, making it suitable for imbalanced datasets. 

\subsubsection{Micro F1-Score}
The micro F1-score \cite{sokolova2009systematic} aggregates the contributions of all classes to compute the average metric. Instead of treating all classes equally, it is weighted by the number of instances in each class, making it more suitable for balanced datasets.

\section{Results}
\label{s:results-discussion}

The results presented in Table \ref{tab:combined_model_comparison} demonstrate the effectiveness and efficiency of the ArabEmoNet architecture for Arabic speech emotion recognition across two distinct datasets: KSUEmotion and KEDAS.

On the KSUEmotion dataset, ArabEmoNet achieves an accuracy of 91.48\%, which represents state-of-the-art performance. This significantly outperforms previously established benchmarks for this dataset, including the CNN-BLSTM-DNN model \cite{hifny2019efficient} and the ResNet-based architecture \cite{meftah2021ksuemotion}. Furthermore, ArabEmoNet also surpasses the performance of larger, pre-trained models such as HuBERT-base \cite{hsu2021hubert} and Whisper-small \cite{radford2022whisper}, despite its significantly smaller parameter count.

Similarly, on the KEDAS dataset, our model achieves an exceptional accuracy of 99.46\%. This result substantially surpasses the original Baseline Model \cite{belhadj2022kedas} and demonstrates competitive performance even when compared to highly resource-intensive pre-trained models like Whisper-small \cite{radford2022whisper} and HuBERT-base \cite{hsu2021hubert}. Notably, ArabEmoNet achieves these superior or competitive results with significantly fewer parameters (0.97M) compared to pretrained models such as HuBERT-base (95M) and Whisper-small (74M).

\section{Discussion and Analysis}

\subsection{CNN Kernel Size}

Table \ref{tab:kernel_size_ablation} shows the impact of kernel size on ArabEmoNet's performance for the KSUEmotion Dataset. As the kernel size increases from 3 to 7, the model's accuracy steadily improves, peaking at 91.48\% with a kernel size of 7 and a corresponding padding of 3. Beyond this point, increasing the kernel size further (to 9 and 11) leads to a decline in accuracy. Larger kernels, while increasing the receptive field, may introduce too much noise or become less adept at capturing fine-grained details, leading to a dip in accuracy. Conversely, smaller kernels might not encompass enough contextual information to achieve optimal recognition. Therefore, the kernel size of 7 represents the best trade-off between performance and model complexity in this experimental setup.

\begin{table}
\centering

\resizebox{0.50\textwidth}{!}{
\begin{tabular}{llcc}
\toprule
\textbf{Kernel Size} & \textbf{Padding} & \textbf{Accuracy (\%)} & \textbf{Params (M)} \\
\midrule
11 & 5 & 89.90 & 1.71 \\
9 & 4 & 91.15 & 1.29 \\
7 & 3 & \textbf{91.48} & 0.97 \\
5 & 2 & 90.08 & 0.71 \\
3 & 1 & 89.71 & 0.55 \\
\bottomrule
\end{tabular}
}
\caption{Impact of Changing Kernel Size for CNN Layers (KSUEmotion Dataset)}
\label{tab:kernel_size_ablation}
\end{table}

\subsection{Data Augmentation}

To assess the contribution of data augmentation to the model's robustness and generalization, we compared the performance of our model trained with and without augmentation techniques on the KSUEmotion dataset. As shown in Table \ref{tab:data_aug_ablation}, employing data augmentation leads to a significant improvement in test accuracy, increasing from 89.10\% to 91.48\%. This improvement demonstrates the effectiveness of data augmentation in enhancing the model's generalization capabilities.

\begin{table}
\centering

\begin{tabular}{lc}
\toprule
\textbf{Training Strategy} & \textbf{Accuracy (\%)} \\
\midrule
Without Augmentation & 89.10\\ 
With Augmentation & \textbf{91.48} \\
\bottomrule
\end{tabular}

\caption{Impact of Data Augmentation on Model Performance (KSUEmotion Dataset)}
\label{tab:data_aug_ablation}
\end{table}

\subsection{Transformer-Based Architecture}

To evaluate different architectural configurations, we performed further experiments with a CNN-Transformer model while keeping the remaining components unchanged. The Transformer-based architecture achieved an accuracy of 86.66\% on the KSUEmotion dataset, as shown in Table \ref{tab:combined_model_comparison}, which is lower than ArabEmoNet's performance of 91.48\%. This comparison suggests that the BiLSTM-based approach is more effective for Arabic dialectical speech emotion recognition tasks.

\section{Conclusion}
\label{s:conclusion}
This study introduces ArabEmoNet, a lightweight yet highly effective architecture for Arabic Speech Emotion Recognition. By integrating 2D CNN layers, BiLSTM networks, and an attention mechanism with Mel spectrogram inputs, ArabEmoNet significantly advances the state-of-the-art, achieving a remarkable 4\% improvement over existing models on the KSUEmotions dataset. Our results demonstrate that 2D convolutions substantially outperform traditional approaches using 1D convolutions and MFCC features, capturing richer and more nuanced acoustic patterns essential for emotion classification.

Furthermore, employing Gaussian noise augmentation successfully enhanced the model's robustness and addressed data imbalance issues, underscoring the importance of effective augmentation strategies. Comparative experiments revealed that transformer-based architectures, while powerful in other contexts, were less effective for this task, highlighting the particular suitability of BiLSTM layers in capturing temporal emotional dynamics.

In future work, we aim to extend ArabEmoNet's training to larger, multilingual datasets, validating its applicability and generalizability across diverse linguistic and cultural contexts. This expansion promises significant contributions toward more inclusive and effective global emotion recognition systems.

\section{Limitations}
A potential limitation to our architecture arises from the method used to handle variable audio lengths. To standardize the input size for model processing, the architecture employs zero-padding. Specifically, shorter audio sequences within any given batch are padded with zeros to equal the length of the longest sequence in that same batch. While this is a standard technique, it can introduce a limitation if there is significant variance in the duration of audio clips within a batch. In such cases, shorter clips will be appended with a large amount of non-informative zero values, which can lead to unnecessary computational processing and potentially impact the model's learning efficiency

\bibliography{custom}

\end{document}